\RequirePackage{lineno}
\documentclass[aps,prl,twocolumn,superscriptaddress,showpacs,floatfix]{revtex4-1}
\usepackage{graphicx}
\usepackage{color,textcomp}
\usepackage{amssymb}

\begin{document}
\linenumbers
\title{Phonon Dispersion and Elastic Moduli of Two-Dimensional Disordered Colloidal Packings of Soft Particles with Frictional Interactions}

\author{Tim Still}
\email[Electronic address: ]{timstill@seas.upenn.edu}
\affiliation{Department of Physics and Astronomy, University of Pennsylvania, Philadelphia PA 19104, USA}
\affiliation{Complex Assemblies of Soft Matter, CNRS-Rhodia-UPenn UMI 3254}
\author{Carl P.\ Goodrich}
\affiliation{Department of Physics and Astronomy, University of Pennsylvania, Philadelphia PA 19104, USA}
\author{Ke Chen}
\affiliation{Beijing National Laboratory for Condensed Matter Physics and Key Laboratory of Soft Matter Physics, Institute of Physics, Chinese Academy of Sciences, Beijing 100190, China}
\author{Peter J.\ Yunker}
\affiliation{Department of Physics and Astronomy, University of Pennsylvania, Philadelphia PA 19104, USA}
\affiliation{School of Engineering and Applied Sciences, Harvard University, Cambridge, Massachusetts, USA}
\author{Samuel Schoenholz}
\author{Andrea J.\ Liu}
\author{A.~G. Yodh}
\affiliation{Department of Physics and Astronomy, University of Pennsylvania, Philadelphia PA 19104, USA}

\date{\today}

\begin{abstract}

Particle tracking and displacement covariance matrix techniques are employed to investigate the phonon dispersion relations of two-dimensional colloidal glasses composed of soft, thermoresponsive microgel particles whose temperature-sensitive size permits \textit{in situ} variation of particle packing fraction. 
Bulk, $B$, and shear, $G$, moduli of the colloidal glasses are extracted from the dispersion relations as a function of packing fraction, and variation of the ratio $G/B$ with packing fraction is found to agree quantitatively with predictions for jammed packings of frictional soft particles.
In addition, $G$ and $B$ individually agree with numerical predictions for frictional particles.  
This remarkable level of agreement enabled us to extract an energy scale for the inter-particle interaction from the individual elastic constants and to derive an approximate estimate for the inter-particle friction coefficient.

\end{abstract}


\maketitle


Like a madeleine dipped in tea, a packing of ideal spheres at the jamming transition is barely solid.  The ratio of the shear modulus to the bulk modulus, $G/B$, vanishes, as it does for a liquid, and the number of inter-particle contacts is exactly the minimum number needed for mechanical stability, namely the isostatic number, $z_c=2D$, where $D$ is the dimensionality~\cite{ohern03}.  
Above the jamming transition, $G/B$ increases linearly with the number of excess contacts, $z-z_c$~\cite{ohern03,vanhecke10,liu10}, or equivalently, with $\left(\phi-\phi_c\right)^{1/2}$, where $\phi$ is the packing fraction and $\phi_c$ is the packing fraction at the transition.  
This scaling relation is a defining property of the jamming transition; it sets jammed packings apart from other solids whose inter-particle contact number can be varied above the isostatic value, such as networks near the rigidity percolation threshold~\cite{ellenbroek09} and decorated isostatic lattices~\cite{souslov09,mao11}.

Despite its central importance to jamming transition theory, the behavior of $G/B$ has proven challenging to measure experimentally.  
Among all the relations predicted near the jamming transition~\cite{liu10}, only the dependence of the excess contact number, $z-z_c$~\cite{katgert10,jorjadze13}, and the bulk modulus, $B$~\cite{jorjadze13}, on the excess packing fraction, $\phi-\phi_c$, have been tested experimentally.  
Here we circumvent traditional technical difficulties in measuring by $G$ and $B$ by employing video microscopy on two-dimensional disordered colloidal packings to measure phonon dispersion relations. 

The glassy colloidal suspensions are composed of poly(N-isopropyl acrylamide) (PNIPAM) soft hydrogel particles, whose packing fraction, $\phi$, can be tuned \textit{in situ} by changing temperature.  
Such systems have proven useful for studying the properties of colloidal packings near the jamming transition~\cite{chen10prl,zhang11,yunker11prl,chen11}.  
We employ displacement covariance matrix analysis \cite{ghosh10prl,chen11,chen10prl,kaya10,henkes11} to obtain the system's eigenmodes and eigenfrequencies.
Using an analysis similar to those in earlier studies~\cite{keim04,kaya10,klix12}, we obtain the phonon dispersion relation, $\omega(q)$, for the vibrational modes of a ``shadow system" with equivalent particle configuration and interactions but without damping.  
Sound velocities and elastic moduli are then derived from the dispersion relation. 

While most studies of colloidal suspensions are interpreted without invoking direct frictional interactions between particles, such effects can arise~ \cite{kurita10,liber13}.
The present experiments allow direct comparison with models of jammed systems with~\cite{somfai07,shundyak07} and without~\cite{ohern03} inter-particle friction. 
We find unambiguously that the particles have frictional interactions; from the dependence of the elastic constants on packing fraction, we extract an estimate of the coefficient of friction, $\mu$, as well as the strength of inter-particle interactions, $\epsilon$.

PNIPAM particles with different diameters were synthesized by surfactant-free radical emulsion polymerization, as described elsewhere \cite{alsayed11}.
These particular PNIPAM particles are more strongly cross-linked in their cores compared to their surfaces, and they are essentially charge neutral; thus, when pressed close together, polymeric chains of one particle are very likely to interpenetrate and entangle with particle chains of neighboring particles.
Quasi-2D packings (binary mixtures with $d_{big}\approx1.4$~\textmu m and $d_{small}\approx1.0$~\textmu m (at 26~$^{\circ}$C)) were prepared by confining the suspension between two microscope cover slips (Fisher Scientific) and sealed with optical glue \cite{han08pre}.
The diameter of the PNIPAM particles changes with temperature, $T$; $d(T)$-curves obtained by dynamic light scattering can be found in the supporting material \cite{supple}.
Particle trajectory data were acquired using standard bright field video microscopy in a narrow range of temperatures, $26.4-27.2$~$^{\circ}$C.
The temperature was controlled by thermal coupling to the microscope objective (BiOptechs), and the sample was equilibrated for 15 min at each temperature before data acquisition.  
During this 15 min period, particle rearrangements occurred as the system aged; to our knowledge, no cage rearrangements occurred once data acquisition was begun, except in the system at 27.2$^{\circ}$C, $\phi\approx0.863$, the lowest $\phi$ studied.
Note that the diameter ratio, $d_{big}/d_{small}$, varies by less than within 1\% in the investigated range of temperatures.

The trajectories of the $N\approx3000$ particles in the field of view ($\approx67\times50$~\textmu m) were extracted from a total of 30,000 frames of video at 110 frames/s using standard particle tracking techniques. 
The packing fraction $\phi$ was calculated from the measured number of particles and their hydrodynamic radii (measured at low concentration) at the experiment temperature.
Note that \textit{changes} in this ``hydrodynamic'' packing fraction accurately reflect changes in the true packing fraction.
The absolute packing fractions, on the other hand, are typically overestimated because the hydrodynamic radius in dynamic light scattering experiments tends to be larger than the diameter measured by static scattering techniques or direct imaging.

To analyze the data, we employ the displacement covariance matrix technique \cite{ghosh10prl,chen10prl,kaya10,henkes11}.  
We define $\mathbf{u}(t)$ as the 2$N$-component vector of displacements of all particles from their time-averaged positions, and we extract the displacement covariance matrix, $C_{ij}=\left\langle u_i(t)u_j(t)\right\rangle_t$,	
where $i$, $j=1\ldots 2N$ run over particles and coordinate directions, and the average is taken over time frames.
In the harmonic approximation and in thermal equilibrium, $C_{ij}$ is directly related to the dynamical matrix of the shadow system
$D_{ij}={k_BTC_{ij}^{-1}}/{\sqrt{m_i m_j}}$ with particle masses $m_{i}$ and $m_{j}$. 
The eigenvectors of $D$ are the vibrational eigenmodes of the shadow system with polarization vectors $\mathbf{P}_n$ (for $n=1\ldots 2N$) and eigenfrequencies $\omega_n=\left(\frac{k_BT}{m\lambda_n}\right)^{1/2}$, where $\lambda$ are the eigenvalues of covariance matrix $C$, and $m$ is the mass of a single sphere.

The vibrational mode frequencies thus extracted depend on the experimental number of snapshots.  
We correct for the error that arises from using a finite number of frames by extrapolating to $N_\mathrm{frames}=\infty$ and assuming that $\omega$ varies linearly in $1/N_\mathrm{frames}$, as expected~\cite{schindler12,chen13}.

\begin{figure}[tb]
	\centering
		\includegraphics[width=.9\linewidth]{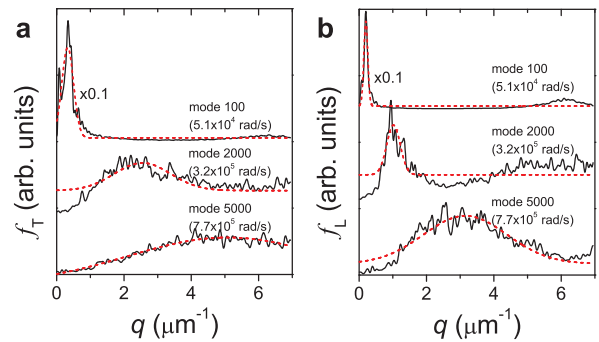}
	\caption{(color online) 
		\textbf{a} Transverse and \textbf{b} longitudinal spectral functions, $f_T(\omega)$ and $f_L(\omega)$, of the system at low, medium, and high $\omega$. 
		Dashed lines show Gaussian fits to the peaks which were used to extract $q(\omega)$ for each $\omega$.}
	\label{fig:1}
\end{figure}

The Fourier decomposition of the eigenmodes into transverse and longitudinal components yields two spectral functions, $f_T$ and $f_L$, respectively, for each mode of frequency $\omega$ as a function of wavevector magnitude $q$:
\begin{eqnarray}
\label{eq:ft}
f_T(q,\omega)&=&\left\langle \left| \sum_n \widehat{\mathbf{q}}\times\mathbf{P}_n(\omega)\exp(i\mathbf{q}\cdot\mathbf{r}_n)\right|^2\right\rangle,\\
f_L(q,\omega)&=&\left\langle \left| \sum_n \widehat{\mathbf{q}}\cdot\mathbf{P}_n(\omega)\exp(i\mathbf{q}\cdot\mathbf{r}_n)\right|^2\right\rangle, 	
\label{eq:fl}
\end{eqnarray}
where $\mathbf{r}_n$ is the equilibrium position of each particle and the brackets indicate an average over directions $\widehat{\mathbf{q}}$ \cite{grest82,silbert09,vitelli10}.  

The maxima of these functions correspond to the phonon wavevector with magnitude $q_{T,L}(\omega)$ that constitute the dispersion relation \cite{silbert09,kaya10}.  
We recently applied this method to a hexagonal colloidal crystal \cite{chen13} and obtained the full dispersion relation expected theoretically~\cite{ashcroft}, as have earlier colloidal experiments \cite{keim04,grunberg07,reinke07,baumgartl08}.   
In the long wavelength limit, the dispersion curve is linear and its slope gives the longitudinal and transverse sound velocities: $c_{T,L}=\lim\limits_{q\rightarrow0}(\partial \omega/\partial q)$.

In practice, the procedure for extracting the maximum value of $f_{T,L}$ as a function of $q$ for each mode yields rather noisy results for disordered colloidal packings~\cite{supple}, as expected from numerical studies of jammed packings \cite{vitelli10}.  
In contrast to crystals, where the peak in $f_{T,L}$ is very sharp, for disordered systems it has been shown~\cite{silbert09} that the peak is relatively broad and flat for frequencies above the so-called ``boson peak frequency'' \cite{silbert05}, $\omega^*$ (which is 30-80$\times10^3$ rad/s for our experimental systems).    
To extract the maximum of $f_{T,L}(q,\omega)$ more cleanly for each mode, we therefore fit $f_{T,L}$ to a Gaussian in $q$ to obtain $q_{max}$.  
Representative plots of $f_{T,L}$ are shown for three different modes in Fig.~\ref{fig:1}, along with the fits used to obtain $q_{max}$ for each mode. 
Since a glass should be isotropic, we average over many ($>100$) directions in Eqs.~\ref{eq:ft}-\ref{eq:fl} to improve the statistics.  

\begin{figure}
	\centering
		\includegraphics[width=1\linewidth]{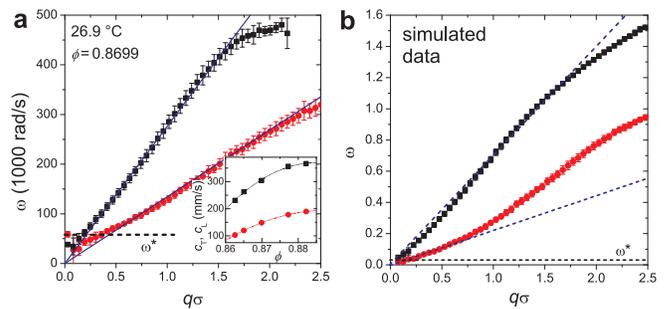}
	\caption{(color online) \textbf{a} Experimental dispersion relation for PNIPAM glass with $\phi\approx0.87$ (squares: longitudinal, circles: transverse) plotted vs.\ $q\sigma$ with average diameter $\sigma\approx1.1$~\textmu m. The horizontal dashed line marks $\omega^\star \approx 60$~rad/s.  Solid lines show linear fits in the long wavelength limit used to extract the sound velocities, $c_L$ and $c_T$.
\textbf{inset:} Sound velocities for the five investigated packing fractions. Second order polynomials (dotted lines) are guides for the eye.
\textbf{b} Numerical dispersion relation for frictionless particles with harmonic repulsions (bidisperse with diameter ratio 1.4 at $p=10^{-2}$). The horizontal dashed line marks $\omega^\star\approx0.03$. Dashed red lines show the slopes that would correspond to the elastic moduli measured directly in the simulation ($G=0.053$ and $B=0.43$).}
	\label{fig:2}
\end{figure}

The resulting dispersion relations are shown in Fig.~\ref{fig:2}a for the intermediate packing fraction. 
The transverse (red circles) and longitudinal (black squares) branches are binned in $q$; the error bars show the standard error of all $\omega$ in the bin.  For all $\phi$, the dispersion relation can be obtained from Gaussian fits at least up to $q\approx2$ \textmu m$^{-1}$.
In all, we studied five different packings in the range $0.8626\le\phi\le0.8822$; the remaining four sets of dispersion curves are shown in the supplementary material~\cite{supple}.
In all cases, the curves are essentially linear at low $q$ and bend at higher $q$, as expected~\cite{silbert09, klix12}.

\begin{figure}
	\centering
		\includegraphics[width=.92\linewidth]{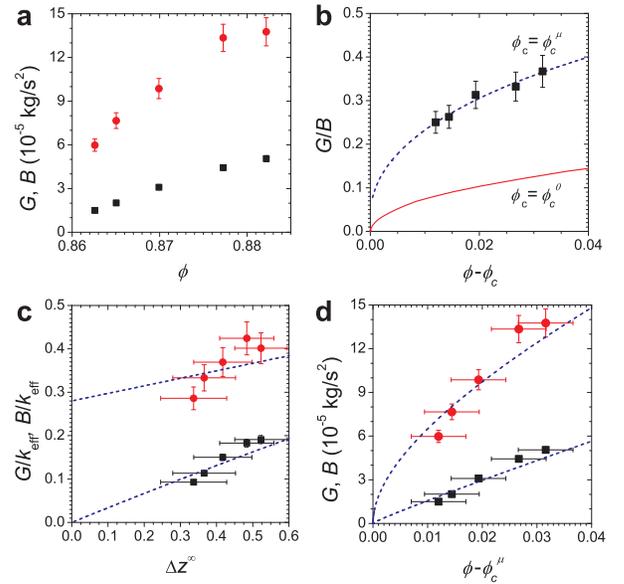}
	\caption{(color online)  
\textbf{a} Experimental bulk ($B$, circles) and shear moduli ($G$, squares) as a function of packing fraction $\phi$.
\textbf{b} Ratio $G/B$ as a function of $\phi-\phi_c$.  Dashed line shows the expected curve for frictional spheres~\cite{somfai07}, where $\phi_c^\infty$ and  $\phi_c^\mu$ are the fit parameters. For comparison, the solid red curve shows $G/B$ calculated for frictionless particles.   
\textbf{c} $B/k_{eff}$ and $G/k_{eff}$ as a function of $\Delta z^\infty=z-z_c^{\infty}$ with corresponding fits (see text); $\epsilon$ is the only fit parameter.
\textbf{d} $B$ and $G$ as a function of $\phi-\phi_c^{\mu}$. Dashed lines are the same fits as in \textbf{c}.}
		\label{fig:3}
\end{figure}

Note that most of the data lie at frequencies above the boson peak frequency, $\omega^*$.  Previous simulations found that while the dispersion relation for $\omega<\omega^*$ is linear in $q$, with a slope consistent with the elastic constant expected for sound modes~\cite{vitelli10}, for $\omega \gg \omega^*$ the situation is different: the modes are not plane-wave-like and the distinction between transverse and longitudinal directions breaks down~\cite{silbert09}. To corroborate that the elastic constants can be extracted from dispersion relations above $\omega^*$, we compare to numerical calculations. 
Fig.~\ref{fig:2}b shows the transverse and longitudinal dispersion relations of numerically-generated jammed bidisperse packings, extracted by fitting the peaks of $f_{T,L}(q,\omega)$; the dispersion relations remain linear in $q$ up to frequencies about an order of magnitude higher than $\omega^*$, with slopes consistent with the values of the elastic moduli, as indicated by the dashed lines. 
These calculations were carried out at  pressure $p=10^{-2}$, where $\omega^* \approx 0.03$ is in units of $\sqrt{\epsilon/m \sigma^2}$, where $\epsilon$ is the interaction strength of the particles which interact via harmonic repulsion, $\sigma$ is the average particle diameter, and $m$ is the particle mass.  These results suggest that in analyzing the data, we must restrict ourselves to a range of frequencies within an order of magnitude of $\omega^*$ in order to extract the sound velocities from linear fits to the experimentally-obtained dispersion relations, i.e., over the range $0.25<q<1.00$ \textmu m$^{-1}$ (solid blue lines). 
The mass density $\varrho$ of the particles and the entire system is very close to that of water (1000 kg/m$^3$), and the areal density is  $\rho_{2D}=\varrho h$, where $h\approx1.4\cdot10^{-6}$~m is the height of the sample cell. We thus obtain the longitudinal modulus, $M=\rho_{2D} c_l^2$, the shear modulus $G=\rho_{2D} c_t^2$, and the bulk modulus, $B=M-G$ for each packing fraction  (Fig.~\ref{fig:3}a).

We also compare the magnitude of $G$ from the experimental dispersion relation to that measured in bulk rheology experiments of jammed PNIPAM suspensions \cite{purnomo08,carrier09,nordstrom10}. 
Expressed in 3D-units, we find $G$ to vary between $\approx 10-36$~Pa, consistent with earlier measurements on similar systems, which found a range $4-20$~Pa~\cite{senff99,purnomo08,carrier09}.

 According to theoretical predictions for athermal systems near the jamming transition~\cite{ohern03,ellenbroek09}, the ratio of the shear to bulk modulus, $G/B$, should be independent of the inter-particle potential.  
For the frictionless case, numerically-generated packings are well-described by $G/B \approx 0.23 \Delta z (1-0.14 \Delta z)$, where $\Delta z=z-z_c^0$ and the \textit{frictionless} isostatic number is $z_c^0=2D=4$ in two dimensions.

By contrast, for frictional particles, we find
\begin{equation}
	G/B =0.8(\pm0.1)\Delta z^\infty (1-0.25(\pm0.05) \Delta z^\infty) 
	\label{eq:gb}
\end{equation}
by fitting simulation data in Fig.~4b of Somfai \textit{et al.}~\cite{somfai07}, where $\Delta z^\infty=z-z_c^\infty$ and the \emph{frictional} isostatic number at \emph{infinite friction} is $z_c^{\infty}=D+1=3$.  

Unfortunately, it is very difficult to deduce the contact number directly from experiment.  
We can, however, analyze the experimental findings using our packing fraction measurements and a result obtained from numerical simulations of frictional particles~\cite{shundyak07}.
For particles with finite friction coefficient $\mu$, the scaling relation between $z-z_c^\infty$ and $\phi-\phi_c^\infty$, where $\phi_c^\infty$ is the critical packing fraction at infinite friction, depends on the critical packing fraction for particles with friction $\mu$, $\phi_c^\mu$ (note, $\phi_c^\infty\le\phi_c^\mu\le\phi_c^0$) \cite{shundyak07}.
Using $z-z_c^\mu=C_1(\phi-\phi_c^\mu)^{0.5}$  and $z_c^\mu-z_c^\infty=C_2(\phi_c^\mu-\phi_c^\infty)^{1.7}$, we fit our data to Eq.~\ref{eq:gb} with $\Delta z^\infty=(z-z_c^\mu)+(z_c^\mu-z_c^\infty)=C_1(\phi-\phi_c^\mu)^{0.5}+C_2(\phi_c^\mu-\phi_c^\infty)^{1.7}$; from Ref.~\cite{shundyak07}, $C_1=2.7\pm0.6$, and $C_2=65\pm2$. 
We note that the fitting involves 
two fit parameters, $\phi_c^\mu$ and $\phi_c^\infty$. 
(Note also, because $\phi_c^\mu-\phi_c^\infty$ is a function of $\mu$, we could have used $\mu$ as the second fit parameter instead of $\phi_c^\mu$.)

The resulting best fit for $G/B$ as a function of $\phi-\phi_{c}$ is shown in Fig.~\ref{fig:3}b (dashed line, $\phi_c=\phi_c^\mu$) together with the expected curve for the frictionless form (solid line, $\phi_c=\phi_c^0$).
The agreement is excellent with the frictional form, whereas the agreement with the frictionless form is poor. 
The results therefore lead us to conclude that PNIPAM particles in suspension experience inter-particle friction effects.    

The fit parameters are $\phi_c^\mu\approx0.851\pm0.005$ and $\phi_c^\infty\approx0.837\pm0.01$, indicating a $\mu$ of order unity or higher by comparison to Ref.\ \cite{shundyak07}.
We note further that $\phi_c^\infty$, and thus the difference $\phi_c^\mu-\phi_c^\infty$, is particularly sensitive to small changes in $C_1$, $C_2$, and to the coefficients in Eq.~\ref{eq:gb}. 
This sensitivity limits the accuracy of our  determination of $\mu$. 
In addition, there is significant uncertainty arises in such estimates, because the relationship between $\mu$ and $z$ is model-dependent \cite{papanikolaou13}.

The inter-particle friction effects suggested by the data above are consistent with expectations based on the structure of the colloidal particles. 
On a molecular scale, the particular hydrogel particles utilized here are hairy at their surfaces, with polymer chains extending freely into the solvent. 
Thus, when the colloidal particles are pushed closely together, entanglement of polymer chains and attractive van-der-Waals interactions can arise and contribute to inter-particle friction.  

In two dimensions, $\phi_c^\infty$ is expected to correspond to the random loose packing fraction, $\phi \approx 0.76$~\cite{silbert10,meyer10,papanikolaou13}.  
The discrepancy with our fitted value of $\phi_c^\infty\approx0.84$ is therefore quite reasonable, given the uncertainties associated with hydrodynamic radius.  
In this spirit, the packing fractions could be corrected from the hydrodynamical derived values by subtracting $\approx0.08$.

We next show that analysis of the individual elastic constants, $G$ and $B$, allows us to extract the interaction energy. 
Previous experiments by Nordstrom \textit{et al.}~\cite{nordstrom10} suggest that the particle interaction potential has the Hertzian form, i.e. $V(r_{ij}) = \frac{\epsilon}{5/2} \left( 1 - r_{ij}/\sigma_{ij} \right)^{5/2}$  for overlap of particles $i$ and $j$, and $V(r_{ij})=0$ otherwise.    
Here, $r_{ij}$ is the center-to-center particle separation, $\sigma_{ij}$ is the sum of their radii, and $\epsilon$ sets the interaction energy scale.  
In previous work, Somfai {\it et al.}~\cite{somfai07} studied the effects of different $\mu$ on the elastic moduli of systems of frictional Hertzian particles in 2D. 
Here we utilize their simulation results to show that our data collapse onto a single curve when $G/k_{eff}$ is plotted against $\Delta z$. 
The same is true for $B/k_{eff}$. 
Here, $k_{eff}=\frac{\sqrt{3}\epsilon}{2\sigma^2} (p/p_0)^{1/3}$ for Hertzian interactions; $p=p_0 (\phi-\phi_c^\mu)^{3/2}$, where $p_0=0.135$ for frictional particles \cite{shundyak07}.
Using this form, the numerical data of Somfai {\it et al.}~\cite{somfai07} are described by $G/k_{eff} \approx  0.34 \Delta z^\infty (1- 0.09\Delta z^\infty)$, and $B/k_{eff} \approx 0.28 (1+ 0.62 \Delta z^\infty)$ \cite{supple}.

In short, with $\epsilon$ as a single fit parameter, we can fit experimental data to the theoretically expected results for $G/k_{eff}$ and $B/k_{eff}$ derived from simulations of Hertzian particles with friction in 2D.   Note that these fits rely on $\phi_c^\infty$ and $\phi_c^\mu$, which were determined previously from the fit to $G/B$, and so they are fixed in this analysis.
The results are shown in Fig.~\ref{fig:3}c; we find $\epsilon\approx6\pm1\times10^5$~$k_BT$.   
Equivalently, we show $G$ and $B$ versus $\phi-\phi_c^\mu$ in Fig.~\ref{fig:3}d.

Thus far, we have examined our experimental system in the context of theoretical predictions for disordered packings at zero temperature.  
Our particles, however, are thermal with $k_B T/\epsilon \approx 2\times 10^{-6}$.  This temperature may seem very low, but recent simulations suggest that thermal effects can dominate even in this range.
For example, it has been suggested that similar experiments with PNIPAM systems ~\cite{chen10prl,chen11} have failed to probe the physics of the jamming transition because $k_B T/\epsilon$ is too high.
Specifically, the simulations of Ikeda {\it et al.}~\cite{ikeda13jcp} on systems with harmonic repulsions suggest that the scaling behavior of the jamming transition is recovered only for temperatures lying below $k_B T_{Ikeda}^\star/\epsilon \approx 10^{-3}(\phi-\phi_c)^2$.   Simulations of Wang and Xu~\cite{wang13} recover jamming scaling for $k_B T_{Wang}^\star/\epsilon \approx 0.2 (\phi-\phi_c)^2$.  
Note that the same scaling with $\phi-\phi_c$ is observed by both Ikeda \textit{et al}. and by Wang and Xu; this scaling is determined by the form of the interaction energy.   
However, the prefactors found by the two groups differ by roughly a factor of 100.  
This difference in prefactors arises because $T^\star$ is a crossover temperature, not a transition temperature.  As a result, it is not well-defined, and the value of the prefactor depends on the measure used.  

For systems with Hertzian repulsions, such as ours, one would expect $k_B T^\star/\epsilon \sim (\phi-\phi_c)^{5/2}$ with a prefactor that is similar to the harmonic case~\cite{wang13}. 
For the lowest packing fraction studied, $\phi-\phi_c^\mu \approx 0.012$, giving $k_B T_{Ikeda}^\star/\epsilon \approx 1 \times 10^{-7}$ and $k_B T_{Wang}^\star/\epsilon \approx 3 \times 10^{-5}$, respectively; in this case our measured value satisfies $T_{Ikeda}^\star < T  \lesssim T_{Wang}^\star$.  
Therefore, we should not recover jamming-like behavior according to Ikeda {\it et al.}, but should be at the border of recovering jamming-like behavior according to Wang and Xu.  
The fact that our results are in quantitative agreement with $T=0$ predictions suggests that the prefactor of Wang and Xu is more consistent with our experimental observations.  

Further evidence that our experiments can be analyzed in terms of the athermal results is provided by the root-mean-squared displacement, $\Delta r$. 
In particular, we find that $\Delta r$ is comparable to the estimated particle-particle overlap at the lowest $\phi$ studied, indicating again that this data point is borderline and is about one order of magnitude smaller than particle-particle overlap at the highest $\phi$ \cite{supple}. 
Thus, our analysis of the data in terms of the zero-temperature theory is justified, with the possible exception of the lowest $\phi$ data point.

To conclude, we have employed colloidal suspensions of temperature-sensitive particles to probe the scaling of the bulk and shear elastic moduli as a function of packing fraction in the vicinity of the jamming transition. 
The observed scaling behaviors are quantitatively consistent with the predictions of jamming theory for frictional particles. 
Our results suggest that static friction is important, at least in the concentrated PNIPAM colloidal packings studied here. 
In granular materials, friction is also important, but thermal effects are negligible; by contrast, for colloidal systems, the interplay of friction and temperature requires exploration.  
To date, these types of systems are typically interpreted using glass theories at nonzero temperatures without friction or jamming theories (with or without friction) in the athermal limit. Our findings suggest that (soft) colloids belong to a sample class wherein thermal effects and friction effects might need to be considered.
In the future it should be possible to manipulate and study such friction effects by changing colloidal particle softness, size, and interaction, as well as to tune from the athermal regime, which describes our results well, to the thermal regime.

We thank R. Kamien, T. Lubensky, K. Ap\-to\-wicz, M.\ van Hecke, M.\ Gratale, and M.\ Lohr for helpful discussions, and gratefully acknowledge financial support from NSF, through the PENN MRSEC DMR11-20901, DMR12-05463, and a Graduate Fellowship (CPG) and from NASA NNX08AO0G.
T.~S. acknowledges financial support from DAAD.


\end{document}